\documentclass[10pt,twocolumn,pre,aps,superscriptaddres,floatfix]{revtex4-1}

\usepackage{amssymb}
\usepackage{amsmath}
\usepackage{bm}% bold math

\usepackage[linkcolor = blue, citecolor = blue, urlcolor = blue, colorlinks = true]{hyperref}
\usepackage{cleveref}

\usepackage{graphicx}
\usepackage{wrapfig}
\usepackage[dvipsnames]{xcolor}

\usepackage{comment}
\usepackage[utf8]{inputenc}
\usepackage{lipsum}

\begin{document}

\title{Hydrodynamic Interactions in Particle Suspensions:\protect\\ A  Perspective on Stokesian Dynamics}

\author{K.W. Torre}
 \email{k.w.torre@uu.nl}
\author{J. de Graaf}
 \email{j.degraaf@uu.nl}
\affiliation{
 Institute for Theoretical Physics, Center for Extreme Matter and Emergent Phenomena, Utrecht University, Princetonplein 5, 3584 CC Utrecht, The Netherlands 
}

\date{\today}

\begin{abstract}
Stokesian Dynamics (SD) is a numerical framework used for simulating hydrodynamic interactions in particle suspensions at low Reynolds number. It combines far-field approximations with near-field lubrication corrections, offering a balance between accuracy and efficiency. This work reviews SD and provides a perspective on future directions for this approach. We outline the mathematical foundations, the method's strengths and weaknesses, and the computational challenges that need to be overcome to work with SD effectively. We also discuss recent advancements that improve the algorithm's efficiency, including the use of iterative solvers and matrix-free approaches. In addition, we highlight the limitations of making stronger, albeit more cost-effective approximations to studying hydrodynamic interactions in dense suspensions than made in SD, such as the two-body Rotne-Prager-Yamakawa (RPY) approximation. To overcome these issues, we propose a hybrid framework that replaces SD's full many-body computations with a neural network trained on SD data. That is, we correct the RPY approximation, while avoiding costly matrix inversions. We demonstrate the potential of this method on a simple system, where we find a close match to SD data while algorithmically outperforming RPY. Our work provides an outlook on the way in which large-scale simulations of particle suspensions can be performed in the foreseeable future.
\end{abstract}

\maketitle

%%%%%%%%
\section{Introduction}
%%%%%%%%

Particle suspensions --- solid particles in a continuous liquid medium~\cite{happel2012low, kim2013microhydrodynamics} --- make up a large fraction of household products~\cite{tadros2014formulation}, play a key role in industrial processes~\cite{russel1991colloidal}, are ubiquitous in rheology and soft-matter research~\cite{larson1999structure}, and are relevant for biological systems~\cite{spagnolie2015complex}. Examples include personal care products such as shampoos, deodorants, and toothpastes~\cite{larson1999structure}, foodstuffs like milk, fruit juices, and soups~\cite{foodstuffs}, as well as biological fluids such as blood~\cite{darras2022}, mucus, and synovial fluid~\cite{spagnolie2015complex}. Typically, the particles in these examples can undergo (slight) deformations. However, for many systems of interest these remain below $1\%$ of the particle diameter, making them practically negligible~\cite{happel2012low, kim2013microhydrodynamics}. Then, particles can be effectively treated as rigid solids without significantly affecting the accuracy of the hydrodynamic model. Abstracting further, it is useful to consider suspensions of spheres with equal diameter,~\textit{i.e.}, monodisperse hard spheres.

The motion of particles relative to the fluid generates dissipative forces that significantly influence the overall dynamics of the suspension~\cite{kim2013microhydrodynamics}. A moving particle not only experiences drag from the fluid but also disturbs the surrounding liquid flow. This disturbance propagates throughout the medium, influencing neighboring particles, which in turn modify the velocity field experienced by others. The fluid-mediated couplings between particles are referred to as hydrodynamic interactions (HIs). These are inherently long-ranged, with the flow disturbance caused by a single translating particle decaying inversely with interparticle distance. Even weakly disturbed fluid regions can exert noticeable effects on nearby particles, leading to correlated motion across the suspension. As a result, particles do not move independently but collectively. This many-body coupling—arising from the interplay between self-induced motion, external forces, and fluid-mediated interactions—is central to the complex collective behavior observed in particle suspensions.

The characterization of HIs has been a longstanding challenge in theoretical and experimental physics, with foundational studies dating back nearly a century~\cite{happel2012low}. The starting point for understanding these interactions are the Navier-Stokes equations~\cite{kim2013microhydrodynamics}. In particle suspensions, dissipation --- proportional to the dynamic viscosity $\mu$ --- dominates over particle inertia --- proportional to the liquid density $\rho$. This is because the small particle size ${a}$ (typically microns) and speed $\Vert \mathbf{U} \Vert$ (microns per second), makes the Reynolds number $\mathrm{Re} = ( \rho U a )/\mu$ for the system very small. It is therefore justified to drop the nonlinear term from the Navier-Stokes equations, reducing the governing equations to the time-dependent Stokes equations. When the unsteady fluid response can also be disregarded, which applies to many soft-matter and granular systems~\cite{happel2012low,kim2013microhydrodynamics}, the fluid response is governed by the Stokes equations. These are a set of linear, steady-state equations, which can be elegantly summarized in two conservation laws
\begin{subequations}
\label{eqn:stokes}
\begin{align}
    \boldsymbol{\nabla} \cdot \boldsymbol{\sigma} &= 0 ; \\
    \boldsymbol{\nabla} \cdot \mathbf{u} &= 0 ,
\end{align}
\end{subequations}
for momentum and volume, respectively. 
Here, $\mathbf{u}$ denotes the fluid velocity field, while the stress tensor is given by
\begin{align} 
\label{eqn:stress} \sigma_{ij} &= - p \delta_{ij} + \frac{\mu}{2} \left( \partial_i u_j + \partial_j u_i \right), 
\end{align}
where $i$ and $j$ index the spatial coordinates, $p$ is the hydrostatic pressure and $\delta_{ij}$ the Kronecker delta.

To obtain a well-posed problem, the fluid velocity is typically required to satisfy no-slip boundary conditions at the surface of each immersed particle, ensuring a unique solution~\cite{ladd1988hydrodynamic}. Additionally, since solving the full boundary-value problem for arbitrarily shaped particles is intractable, the spherical-particle approximation is commonly adopted as a necessary simplification. While real-life suspensions often contain irregularly shaped particles~\cite{spagnolie2015complex}, assuming spheres allows for analytical and numerical treatments that capture key qualitative behaviors~\cite{gelhydroJoost, de2023hydrodynamic, torre2023hydrodynamic}. Even with these simplifications, exact analytical solutions remain limited, with fully resolved results largely restricted to one- and two-sphere problems~\cite{happel2012low, kim2013microhydrodynamics}. For many-particle systems, which are challenging to describe analytically even making significant approximations, numerical approaches have been developed over the last decades to handle many-body HIs~\cite{bolintineanu2014particle}.

The Stokesian Dynamics (SD) method, introduced by Brady and Bossis~\cite{brady1988stokesian}, is a numerical approach for simulating HIs in low Reynolds number systems, which makes use of fundamental solutions. By combining far-field Green's functions with near-field lubrication corrections, SD efficiently models the many-body HIs between particles~\cite{durlofsky1987dynamic}, offering a good balance of computational efficiency and accuracy. However, despite its strengths, SD never saw widespread adoption, due to its overall algorithmic complexity and the computational expense of propagating the particle dynamics. These aspects made it less accessible compared to, for example, lattice-Boltzmann (LB)~\cite{chen1998lattice, ladd2001lattice, dunweg2009lattice, aidun2010lattice, rohm2012lattice}, multi-particle collision dynamics (MPCD)~\cite{malevanets1999mesoscopic, gompper2009multi, howard2018efficient}, and Fluid Particle Dynamics (FPD)~\cite{tanaka2000simulation}. These methods gained popularity due to their versatility in handling complex geometries, thermal fluctuations, and capturing inertial effects. Nevertheless, these approaches struggle to achieve the same level of precision as SD both in the far- and near-field~\cite{bolintineanu2014particle}.

Recent advancements~\cite{sierou2001accelerated, banchio2003accelerated, wang2016spectral, fiore2019fast} have addressed SD's historical bottlenecks, significantly improving its computational performance. Iterative solvers, preconditioning strategies, and matrix-free techniques have also extended SD’s ability toward increasingly dense systems. Fiore and Swan’s Fast Stokesian Dynamics (FSD)~\cite{fiore2019fast} represents a major breakthrough, achieving linear scaling $\mathcal{O}(N)$ by incorporating spectral Ewald methods and optimized matrix-free algorithms. Thus, SD has become a competitive option for simulating HIs, particularly in bulk fluids and systems with simple boundary conditions.

In this paper, we provide a comprehensive review of Stokesian Dynamics, outlining its mathematical foundations, computational methods, and limitations. In Section II, we derive the equations governing HIs between particles, and introduce the mobility and resistance formalisms that are at the basis of SD. We also discuss far-field approximations using multipole expansions and near-field corrections based on lubrication theory. Section III focuses on the equations of motion for colloidal particles, describing the overdamped dynamics and the numerical implementation of SD. In Section IV, we contrast dynamics obtained using SD with that resulting from the Rotne-Prager-Yamakawa (RPY) approximation. We highlight the role of many-body effects and the trade-offs between computational efficiency and accuracy in our comparison. We also discuss recent advancements in fast SD algorithms that improve scalability. This allows us to propose a hybrid framework that leverages exact two-body hydrodynamics while using a neural network to capture many-body effects. We demonstrate its potential for effcient simulations using a simple system. In Section V, we conclude by summarizing SD’s key strengths and outlining future research directions.

%%%%%%%%
\section{Hydrodynamic Coupling} \label{sec:II}
%%%%%%%%

In this section, we describe how identical spherical particles interact \textit{via} the surrounding fluid in overdamped conditions. We start by introducing Fax{\'e}n’s laws. These laws form the basis for defining the hydrodynamic grand mobility and resistance matrices, which provide a systematic way to describe particle-fluid interactions~\cite{happel2012low, kim2013microhydrodynamics}.

Next, we present a far-field approximation for the grand mobility matrix that relies on multipole expansions and which is effective for widely separated particles~\cite{durlofsky1987dynamic}. For particles that are closer together, we approximate the grand resistance matrix using twin-multipole expansions, which improve accuracy for near-field interactions~\cite{jeffrey1984calculation, jeffrey1992calculation}. However, the latter approach neglects many-body effects and is limited to open boundary conditions. These methods together offer a structured way to model hydrodynamic coupling in particle systems~\cite{brady1988stokesian}.

%%%%%%%%
\subsection{Stokes' Equation and Fax{\'e}n's Laws}
%%%%%%%%

We consider $N$ neutrally buoyant spherical particles of radius $a$, suspended in a Newtonian fluid with dynamic viscosity $\mu$. We assume viscous forces to dominate over inertial ones, such that the Reynolds number $\mathrm{Re} \ll 1$. In this regime, the behavior of the fluid is governed by the Stokes equations~\eqref{eqn:stokes}. To connect the linear velocity $\mathbf{U}^{\alpha}$ and angular velocity $\boldsymbol{\Omega}^{\alpha}$  of particle $\alpha$ to the hydrodynamic force $\mathbf{F}^{H,\alpha}$, torque $\mathbf{T}^{H,\alpha}$, and stress $\mathbf{S}^{\alpha}$ acting on it, we use Fax{\'e}n's laws for spherical particles~\cite{faxen}. These describe how the motion of the fluid causes particle movement through imposed no-slip boundary conditions:
\begin{subequations}
 \label{eqn:faxen}
\begin{align}
    \label{eqn:faxen1}
    U^{\alpha}_i - u^{\infty}_i(\mathbf{r}^{\alpha}) &= - \frac{ \ F_i^{\text{H},\alpha}}{6 \pi \eta a } + \left( 1+\frac{a^2}{6} \nabla^2 \right ) u'_i(\mathbf{r}^{\alpha}) \\
    \label{eqn:faxen2}
    \Omega_i^{\alpha} - \omega^{\infty}_i(\mathbf{r}^{\alpha}) &= - \frac{ \ T_i^{\text{H},\alpha}}{8 \pi \eta a^3 } + \frac{1}{2} \epsilon_{ijk} \nabla_j  u'_k(\mathbf{r}^{\alpha}) \\
    \label{eqn:faxen3}
    - E^{\infty}_{ij}(\mathbf{r}^{\alpha}) &= \frac{ 3 S_{ij}^{\alpha}}{{20} \pi \eta a^3 } + \left( 1+\frac{a^2}{10} \nabla^2 \right) e'_{ij}(\mathbf{r}^{\alpha} ).
\end{align}
\end{subequations}
Here, the indices $i$ and $j$ (as well as $k$) represent Cartesian components in three-dimensional (3D) space $\{x, y, z \}$ as before. Summation over repeated indices is implied. The symbol $\epsilon_{ijk}$ in \cref{eqn:faxen2} represents the Levi-Civita permutation tensor, which encodes the antisymmetric properties of the cross product. The term $\mathbf{u'}$ represents the disturbance flow field caused by all particles except the particle of interest (indexed $\alpha$), with the background flow $\mathbf{u}^{\infty}$ subtracted (\textit{i.e.}, the flow in the absence of particles). The term $\mathbf{e'} = 1/2 [ \boldsymbol{\nabla} \otimes \mathbf{u'}+ (\boldsymbol{\nabla} \otimes \mathbf{u'} )^{T} ]$ represents the strain rate of this disturbance flow. Additionally, $\boldsymbol{\omega}^{\infty}$, and $\mathbf{E}^{\infty}$ are the vorticity and strain rate of the background flow, respectively. Notably, in \cref{eqn:faxen3}, there is no kinematic counterpart --- the term $E_{ij}$ is not present ---  because we have assumed that the particles cannot deform, meaning they cannot experience strain~\cite{jeffrey1984calculation, jeffrey1992calculation}.

%%%%%%%%
\subsection{Grand Mobility and Grand Resistance}
%%%%%%%%

The grand mobility matrix is a $11N$-dimensional, symmetric, and positive-definite matrix~\cite{kim2013microhydrodynamics}. It maps hydrodynamic forces (incl. torques) and stresslets to particle velocities (incl. angular ones) and strain rates. Specifically, the Fax{\'e}n relations in \cref{eqn:faxen} can be recast in the form of a matrix-vector product, where the HIs between particles are expressed through the mobility matrix~\cite{durlofsky1987dynamic}. The latter is defined as
\begin{align}
\label{eqn:granmobility_def}
\begin{bmatrix}
\mathcal{U} - \mathcal{U}^{\infty} \\
-\mathcal{E}^{\infty} 
\end{bmatrix}
= - {\mathcal{M}} \cdot 
\begin{bmatrix}
\mathcal{F}^H \\
\mathcal{S}
\end{bmatrix},
\end{align}
where $\mathcal{U}$ is a $6N$-dimensional generalized velocity vector, containing both linear and angular velocities of the particles, and $\mathcal{U}^{\infty}$ represents their counterparts in the imposed flow. The generalized force vector $\mathcal{F}^H$ combines hydrodynamic forces and torques, while $\mathcal{E}^{\infty}$ and $\mathcal{S}$ are $5N$-dimensional vectors that describe the independent degrees of freedom of the imposed rate of strain and the stresslets, respectively. Both $\mathcal{E}^{\infty}$ and $\mathcal{S}$ are derived from symmetric, traceless $3N$-dimensional matrices~\cite{kim2013microhydrodynamics}.

The structure of $\mathcal{M}$ reflects the coupling between kinematic and dynamical degrees of freedom:
\begin{align}
    {\mathcal{M}} = \begin{bmatrix}
    \mathbf{M}_{\mathcal{U}\mathcal{F}} & \mathbf{M}_{\mathcal{U}\mathcal{S}} \\
    \mathbf{M}_{\mathcal{E}\mathcal{F}} & \mathbf{M}_{\mathcal{E}\mathcal{S}} 
\end{bmatrix},
\end{align}
where each sub-block represents a specific interaction. For example, $\mathbf{M}_{\mathcal{U}\mathcal{F}}$ captures how forces and torques affect velocities, while $\mathbf{M}_{\mathcal{E}\mathcal{S}}$ describes the coupling between stresslets and strain rates.

We can similarly define the grand resistance matrix, $\mathcal{R}$, which maps generalized particle velocities to forces and stresslets~\cite{kim2013microhydrodynamics}. The fact that the grand mobility matrix $\mathcal{M}$ is symmetric and positive definite ensures that Eq.~\eqref{eqn:granmobility_def} can be inverted, leading to the definition of the resistance matrix as its inverse:
\begin{align}
\label{eqn:granres_def}
\begin{bmatrix}
\mathcal{F}^H \\
\mathcal{S}
\end{bmatrix}
= - {\mathcal{R}} \cdot 
\begin{bmatrix}
\mathcal{U} - \mathcal{U}^{\infty} \\
-\mathcal{E}^{\infty} 
\end{bmatrix},
\end{align}
where $\mathcal{R} = \mathcal{M}^{-1}$. Physically, the positive definiteness of $\mathcal{M}$ reflects the dissipative nature of viscous flow, ensuring that energy is irreversibly lost due to viscous dissipation~\cite{kim2013microhydrodynamics}. Similar to the grand mobility matrix, $\mathcal{R}$ can be partitioned into sub-blocks that represent specific interactions, resulting in the structure:
\begin{align}
    {\mathcal{R}} = \begin{bmatrix}
    \mathbf{R}_{\mathcal{F}\mathcal{U}} & \mathbf{R}_{\mathcal{F}\mathcal{E}} \\
    \mathbf{R}_{\mathcal{S}\mathcal{U}} & \mathbf{R}_{\mathcal{S}\mathcal{E}} 
\end{bmatrix}.
\end{align}
Each sub-block of $\mathcal{R}$ corresponds to a particular coupling between the generalized velocities and the resulting forces or stresslets. This structure mirrors that of the grand mobility matrix, reflecting the symmetry in their roles within the framework of particle-fluid interactions~\cite{happel2012low, kim2013microhydrodynamics}.
%%%%%%%%
\subsection{Far-Field Interactions}
%%%%%%%%

At any point $\mathbf{r}$ in the fluid, the velocity field $\mathbf{u}$, governed by Stokes' equations \eqref{eqn:stokes}, can be expressed as a sum of Green's functions $\boldsymbol{J}$ integrated over the surface of each particle~\cite{ladyzhenskaya1969mathematical}:
\begin{align}
    \label{eqn:stokes_sol}
    \mathbf{u}(\mathbf{r}) = \mathbf{u}^{\infty}(\mathbf{r}) - \frac{1}{8 \pi \eta}  \sum_{\alpha=1}^N \int_{S_{\alpha}} dS_{\mathbf{r'}} \  \boldsymbol{J}(\mathbf{r}-\mathbf{r'}) \cdot \boldsymbol{f}(\mathbf{r'}).
\end{align}
Here, $S_{\alpha}$ is the surface of particle $\alpha$, $\boldsymbol{f} = \boldsymbol{\sigma} \cdot \mathbf{n}$ the fluid traction, and $\mathbf{n}$ the unit vector normal to the surface. The specific form of the Green's function $\boldsymbol{J}$, often referred to Oseen's tensor in this context, depends on the system's outer boundaries (\textit{e.g.}, periodic, open, or closed)~\cite{kim2013microhydrodynamics, durlofsky1987dynamic, fiore2019fast}. For generality, we leave $\boldsymbol{J}$ unspecified at this stage.

To approximate the velocity field near each particle, we expand $\boldsymbol{J}$ about the particle center $\mathbf{r}^{\alpha}$, defining hydrodynamic moments~\cite{durlofsky1987dynamic} as:
%\begin{align}
%    \label{eqn:hydromoments}
%    \mathbf{Q}^n_\alpha = \int_{S_{\alpha}} dS_{\mathbf{r'}} \ \left [ \prod_{\alpha=1}^{n} \otimes \boldsymbol{\Delta r}^{\alpha} \right ] \otimes \boldsymbol{f}(\mathbf{r'}),
%\end{align}
\begin{align}
\label{eqn:hydromoments}
    Q^{n}_{\alpha, i_1 i_2 \dots i_n} = \int_{S_{\alpha}} dS_{\mathbf{r'}} \, \Delta r^{\alpha}_{i_1} \Delta r^{\alpha}_{i_2} \dots \Delta r^{\alpha}_{i_n} f_j (\mathbf{r'})
\end{align}
where $\boldsymbol{\Delta r}^{\alpha} = \mathbf{r'} - \mathbf{r}^{\alpha}$. The 0th-order moment ($\mathbf{Q}^0_\alpha$) corresponds to the hydrodynamic drag force $\mathbf{F}^{\text{H},\alpha}$, while the first-order moment ($\mathbf{Q}^1_\alpha$) contains antisymmetric (torque) and symmetric (stresslet) components:
\begin{subequations}
\begin{align}
    \label{eqn:monopole}
    {F}^{\text{H},\alpha}_i &= \int_{S_{\alpha}} dS_{\mathbf{r'}} f_{i}(\mathbf{r'}) = Q^{0,\alpha}_{i}, \\
    \label{eqn:dipole_torques}
    T^{\text{H},\alpha}_i &= \epsilon_{ijk} \int_{S_{\alpha}} dS_{\mathbf{r'}} \Delta r_j^{\alpha} f_k(\mathbf{r'}) \nonumber \\
    &= \frac{1}{2} \epsilon_{ijk} (Q^{1,\alpha}_{ij} - Q^{1,\alpha}_{ji}), \\
    \label{eqn:dipole_stresslet}
    {S}^\alpha_{ij} &= \frac{1}{2} \int_{S_{\alpha}} dS_{\mathbf{r'}} [\Delta r'_i f_j(\mathbf{r'}) + \Delta r'_j f_i(\mathbf{r'})] \nonumber \\
    &=  \frac{1}{2}(Q^{1,\alpha}_{ij} + Q^{1,\alpha}_{ji}).
\end{align}
\end{subequations}

Truncating the expansion after the dipole terms and incorporating isotropic corrections from the quadrupole and octupole --- ensuring consistency with the known single-particle solution and capturing essential finite-size effects in the suspension~\cite{durlofsky1987dynamic, kim2013microhydrodynamics} --- the velocity field is approximated as:
\begin{widetext}
\begin{align}
    \label{eqn:velfield1}
    u_i(\mathbf{r}) - u^{\infty}_i(\mathbf{r}) &\approx -\frac{1}{8 \pi \eta} \sum^N_{\alpha=1} \left [ J_{ij}(\mathbf{r}-\mathbf{r}^{\alpha}) \ Q_j^{0,\alpha} + \nabla_k J_{ij}(\mathbf{r}-\mathbf{r}^{\alpha}) \ Q_{kj}^{1,\alpha} + \frac{ \ a^2}{6} \nabla^2 J_{ij}(\mathbf{r}-\mathbf{r}^{\alpha}) F^{\alpha}_i + \frac{ \ a^2}{10}  \nabla^2 \nabla_k J_{ij}(\mathbf{r}-\mathbf{r}^{\alpha}) S^{\alpha}_{jk} \right ] 
\end{align}
The last two terms arise due to the finite size of the particles. They are derived from the case of a single sphere in an infinite fluid, either translating or undergoing pure straining motion~\cite{durlofsky1987dynamic, happel2012low, kim2013microhydrodynamics}. Regrouping the terms we get:
\begin{align}
    \label{eqn:velfield2}
    u_i(\mathbf{r}) &- u^{\infty}_i(\mathbf{r}) \approx - \frac{1}{8 \pi \eta} \sum^N_{\alpha=1} \left [ \left ( 1+\frac{a}{6} \nabla^2 \right ) J_{ij}(\mathbf{r}-\mathbf{r}^{\alpha})  F_j^{\text{H},\alpha} + R_{ij} T^{\text{H},\alpha}_j  
    + \left( 1 + \frac{ \ a^2}{10}  \nabla^2 \right ) \nabla_k J_{ij}(\mathbf{r}-\mathbf{r}^{\alpha}) S^{\alpha}_{jk} \right ].
\end{align}
Here, $R_{ij} = - \epsilon_{ijk} (\nabla_k J_{il} - \nabla_l J_{ik})/4$ is the propagator for the torque, commonly referred to as the ``rotlet''~\cite{durlofsky1987dynamic}. This expresses the velocity field in terms of the forces, torques, and stresslets experienced by the particles.
\end{widetext}

Substituting this velocity field into Fax{\'e}n's laws~\eqref{eqn:faxen1}-\eqref{eqn:faxen3} yields the far-field approximation of the grand mobility matrix, $\mathcal{M}^{\text{ff}}$:
\begin{align}
\label{eqn:granmobility_def}
\begin{bmatrix}
\mathcal{U} - \mathcal{U}^{\infty} \\
-\mathcal{E}^{\infty} 
\end{bmatrix}
\approx - {\mathcal{M}^{\text{ff}}} \cdot 
\begin{bmatrix}
\mathcal{F}^H \\
\mathcal{S}
\end{bmatrix}.
\end{align}
This procedure breaks the problem into pairwise elements, each of which has the same structure. Thus, the sub-blocks of $\mathcal{M}^{\mathrm{ff}}$ can be assembled as:
\begin{align}
{\mathbf{M}^{\text{ff}}_{\mathcal{U}\mathcal{F}}} 
= 
\begin{bmatrix}
\mathbf{M}^{11}_{\mathcal{U}\mathcal{F}} & \mathbf{M}^{12}_{\mathcal{U}\mathcal{F}} & \dots & \mathbf{M}^{1N}_{\mathcal{U}\mathcal{F}} \\[0.5em]
\mathbf{M}^{21}_{\mathcal{U}\mathcal{F}} & \mathbf{M}^{22}_{\mathcal{U}\mathcal{F}} &  & \mathbf{M}^{2N}_{\mathcal{U}\mathcal{F}} \\[0.5em]
\vdots & & \ddots & \vdots\\[0.5em]
\mathbf{M}^{N1}_{\mathcal{U}\mathcal{F}} & \mathbf{M}^{N2}_{\mathcal{U}\mathcal{F}} & \dots & \mathbf{M}^{NN}_{\mathcal{U}\mathcal{F}}
\end{bmatrix},
\end{align}
and similarly for $\mathbf{M}^{\text{ff}}_{\mathcal{U}\mathcal{S}}$, $\mathbf{M}^{\text{ff}}_{\mathcal{E}\mathcal{F}}$, and $\mathbf{M}^{\text{ff}}_{\mathcal{E}\mathcal{S}}$. The components of each sub-block are determined by the Green’s functions and their derivatives evaluated at the center-to-center distances between particles.

Although this approximation captures long-range interactions effectively, it is less accurate for closely spaced particles. This is because the Green's functions decay with inverse distance and the expansion is truncated at the dipole level. Including additional terms would improve accuracy but also increase the matrix dimensionality~\cite{durlofsky1987dynamic}. If truncation occurs before the dipole term, the resulting mobility matrix simplifies to the Rotne-Prager-Yamakawa (RPY) approximation~\cite{rotne1969variational, swan2016rapid, fiore2017rapid}, which maps forces to velocities and excludes torques and stresslets.

Finally, the explicit form of the components in $\mathcal{M}^{\text{ff}}$ depends on the system boundary conditions, whether open~\cite{durlofsky1987dynamic}, closed~\cite{kim2013microhydrodynamics}, or periodic~\cite{fiore2019fast}. Detailed calculations for specific cases are provided in the respective references. For completeness, Appendix~\ref{sec:appen_mobility} includes partial derivations for the case of an infinite fluid. Once the boundary conditions are specified, each sub-block of the grand mobility can be computed for all particle pairs.

%%%%%%%%
\subsection{Near-Field Interactions}
%%%%%%%%

When particles are in close proximity, the multipole expansion used in far-field approximations converges very slowly, requiring numerous terms to achieve acceptable accuracy~\cite{durlofsky1987dynamic}. To address this limitation, near-field interactions are described using the hydrodynamic resistance framework~\cite{kim2013microhydrodynamics, durlofsky1987dynamic, brady1988stokesian, jeffrey1984calculation, jeffrey1992calculation}. Unlike $\mathcal{M}$, which maps forces to velocities, using the grand resistance matrix $\mathcal{R}$ is particularly advantageous at small particle separations. This is because $\mathcal{R}$ can be constructed from analytical two-body expressions~\cite{jeffrey1984calculation, jeffrey1992calculation}, which accurately capture near-field hydrodynamic effects while neglecting higher-order many-body contributions. In this section, we first demonstrate how to compute the two-body resistance and then use these pairwise components to assemble an approximate $N$-body grand resistance matrix.

To compute the components of $\mathcal{R}$  several routes can be taken, including the use of reflections, bispherical coordinates, tangent-sphere and asymptotic expansions~\cite{happel2012low, kim2013microhydrodynamics}. In this work, we follow the twin multipole expansion approach introduced by Jeffrey and Onishi~\cite{jeffrey1984calculation, jeffrey1992calculation}. Although their original publications contained inaccuracies, these have been corrected and expanded upon in successive work~\cite{jeffrey1992calculation}, as well as in Townsend's recent review~\cite{townsend2023generating}.

The resistance framework leverages the linearity of Stokes' equations to express the two-particle resistance relation, from equation~\eqref{eqn:granres_def}, as:
\begin{align}
\label{eqn:res_tensors_def2}
\begin{bmatrix}
\mathcal{F}^H_1 \\[0.5em]
\mathcal{F}^H_2 \\[0.5em]
\mathcal{S}_1 \\[0.5em]
\mathcal{S}_2 
\end{bmatrix}
= - \begin{bmatrix}
\mathbf{R}_{\mathcal{F}\mathcal{U}}^{11} & \mathbf{R}_{\mathcal{F}\mathcal{U}}^{12} & \mathbf{R}_{\mathcal{F}\mathcal{E}}^{11} & \mathbf{R}_{\mathcal{F}\mathcal{E}}^{12} \\[0.5em]
\mathbf{R}_{\mathcal{F}\mathcal{U}}^{21} & \mathbf{R}_{\mathcal{F}\mathcal{U}}^{22} & \mathbf{R}_{\mathcal{F}\mathcal{E}}^{21} & \mathbf{R}_{\mathcal{F}\mathcal{E}}^{22} \\[0.5em]
\mathbf{R}_{\mathcal{S}\mathcal{U}}^{11} & \mathbf{R}_{\mathcal{S}\mathcal{U}}^{12} & \mathbf{R}_{\mathcal{S}\mathcal{E}}^{11} & \mathbf{R}_{\mathcal{S}\mathcal{E}}^{12} \\[0.5em]
\mathbf{R}_{\mathcal{S}\mathcal{U}}^{21} & \mathbf{R}_{\mathcal{S}\mathcal{U}}^{22} & \mathbf{R}_{\mathcal{S}\mathcal{E}}^{21} & \mathbf{R}_{\mathcal{S}\mathcal{E}}^{22} 
\end{bmatrix}
\begin{bmatrix}
\mathcal{U}_1 - \mathcal{U}^{\infty} \\[0.5em]
\mathcal{U}_2 - \mathcal{U}^{\infty} \\[0.5em]
-\mathcal{E}^{\infty}  \\[0.5em]
-\mathcal{E}^{\infty}  
\end{bmatrix}.
\end{align}
Each sub-block of the pair-resistance matrix, following the notation used by Jeffrey and Onishi~\cite{jeffrey1984calculation,jeffrey1992calculation}, is further decomposed as:
\begin{subequations}
\begin{align}
\label{eqn:res_tensors}
    \mathbf{R}_{\mathcal{F}\mathcal{U}}^{\alpha \beta} &= 
    \begin{bmatrix}
    \mathbf{A}_{\alpha \beta} & \mathbf{\Tilde{B}}_{\alpha \beta} \\
    \mathbf{B}_{\alpha \beta} & \mathbf{C}_{\alpha \beta}  
    \end{bmatrix}, \ \ \ \
    \mathbf{R}_{\mathcal{F}\mathcal{E}}^{\alpha \beta} = 
    \begin{bmatrix}
    \mathbf{\Tilde{G}}_{\alpha \beta} \\ 
    \mathbf{\Tilde{H}}_{\alpha \beta}  
    \end{bmatrix}, \\
    \label{eqn:res_tensors2}
    \mathbf{R}_{\mathcal{S}\mathcal{U}}^{\alpha \beta} &= 
    \begin{bmatrix}
    \mathbf{G}_{\alpha \beta} & \mathbf{H}_{\alpha \beta}
    \end{bmatrix}, \ \ \ \
    \mathbf{R}_{\mathcal{S}\mathcal{E}}^{\alpha \beta} = \mathbf{M}_{\alpha \beta}, 
\end{align}    
\end{subequations}
where $\alpha$, $\beta \in \{1,2\}$. The symmetry of $\mathcal{R}$ and the axisymmetric configuration of the two particles about their center-to-center vector $\boldsymbol{\Delta r}$ allow all components to be expressed in terms of 20 scalar functions of $\Delta r$. These scalar functions are derived by solving \textit{ad hoc} resistance problems using twin multipole expansions~\cite{jeffrey1984calculation, jeffrey1992calculation, townsend2023generating}.

We demonstrate the principle by taking two sets of spherical polar coordinates $(r_\alpha, \theta_\alpha, \phi)$, with origins at the centers of the two spheres. Using Lamb's general solution~\cite{happel2012low}, the fluid flow and pressure are expressed as the sum of contributions from each sphere: $\mathbf{u}=\mathbf{u}_1 + \mathbf{u}_2$, and $p= p_1 + p_2$. These contributions are expanded in spherical harmonics as follows:
\begin{subequations}
\begin{align}
    \label{eqn:lambsol}
    p_{\alpha} &= \frac{\eta}{a} \sum_{\substack{m=0 \\ n=m}}^{\infty} p^{(\alpha)}_{mn} \left ( \frac{a}{r_{\alpha}} \right )^{n+1} Y^{m}_{n}(\theta_{\alpha} , \phi), \\ 
    \label{eqn:lambsol2}
    \mathbf{u}_{\alpha} &= \sum_{\substack{m=0 \\ n=m}}^{\infty} \boldsymbol{\nabla} \otimes \left [ q^{(\alpha)}_{mn} \left ( \frac{a}{r_{\alpha}} \right )^{n+1} Y^{m}_{n}(\theta_{\alpha} , \phi) \ \mathbf{r}_{\alpha}  \right ] \nonumber \\ 
    &\qquad + a \ \boldsymbol{\nabla} \left [ v^{(\alpha)}_{mn} \left ( \frac{a}{r_{\alpha}} \right )^{n+1} Y^{m}_{n}(\theta_{\alpha} , \phi) \right ] \nonumber \\
    &\qquad - \frac{n-2}{2n(2n-1)a} r_{\alpha}^2  \ \boldsymbol{\nabla} \left [ p^{(\alpha)}_{mn} \left ( \frac{a}{r_{\alpha}} \right )^{n+1} Y^{m}_{n}(\theta_{\alpha} , \phi) \right ] \nonumber \\ 
    &\qquad + \frac{n+1}{n(2n-1)a} \ p^{(\alpha)}_{mn}  \left ( \frac{a}{r_{\alpha}} \right )^{n+1} Y^{m}_{n}(\theta_{\alpha} , \phi) \ \mathbf{r}_{\alpha}, 
\end{align}    
\end{subequations}
where $Y^{m}_{n}(\theta,\phi)=P^m_n(\cos{\theta}) e^{\mathsf{i}m\phi}$ are spherical harmonics, and $P_n^m$ are associated Legendre polynomials.  Here, $\mathsf{i}$ denotes the imaginary unit, satisfying $\mathsf{i}^2 = -1$, and the symbol $\otimes$ denotes the dyadic (tensor) product, which forms a higher-rank tensor by taking the outer product of two vectors or tensors. Note that, the expansion coefficients $p^{\alpha}_{mn}$, $q^{\alpha}_{mn}$, and $v^{\alpha}_{mn}$ depend only on the sphere separation $r_{\alpha}$ reduced by the radius $a$.

Using these expansions, we study the hydrodynamic problem in a Cartesian coordinate system $\{ \boldsymbol{\hat{x}},\boldsymbol{\hat{y}},\boldsymbol{\hat{z}} \}$, where $\boldsymbol{\hat{z}} = \boldsymbol{\Delta r} / {\Delta r}$, and the $x$-axis is chosen in the plane given by $\phi=0$. The total force, torque, and stresslet exerted by the fluid on each sphere can be computed using equations~\cref{eqn:monopole,eqn:dipole_torques,eqn:dipole_stresslet} and the known properties of spherical harmonics, which simplify surface integrals of harmonic functions to well-established results~\cite{jeffrey1984calculation, jeffrey1992calculation, happel2012low}. The resulting expressions depend exclusively on the expansion coefficients: %(HappelandBrenner section 3.2)
\begin{subequations}   
\begin{align}
\label{eqn:forces_from_twinmultipole}
\mathbf{F}_{\alpha}^{\text{H}} &= -4\pi \eta a \left [ p^{(\alpha)}_{01}(-1)^{3-\alpha} \boldsymbol{\hat{z}} - p^{(\alpha)}_{11} (\boldsymbol{\hat{x}} + \mathsf{i} \boldsymbol{\hat{y}}) \right ], \\
\label{eqn:forces_from_twinmultipole2} \mathbf{T}_{\alpha}^{\text{H}} &=- 8 \pi \eta \left [ q^{(\alpha)}_{01}(-1)^{3-\alpha} \boldsymbol{\hat{z}} - q^{(\alpha)}_{11} (\boldsymbol{\hat{x}} + \mathsf{i} \boldsymbol{\hat{y}}) \right ], \\
\label{eqn:forces_from_twinmultipole3} \mathbf{S}_{\alpha} &= -2 \pi \eta a^2 \Big[ p^{\alpha}_{02} (\boldsymbol{\hat{z}}\otimes\boldsymbol{\hat{z}} - \frac{1}{3} \mathbf{I} ) . \nonumber \\ 
&\qquad - p^{(\alpha)}_{12} [\boldsymbol{\hat{x}} \otimes \boldsymbol{\hat{z}} + \boldsymbol{\hat{z}} \otimes \boldsymbol{\hat{x}} +\mathsf{i}(\boldsymbol{\hat{y}} \otimes \boldsymbol{\hat{z}}+\boldsymbol{\hat{z}} \otimes \boldsymbol{\hat{y}}) ] 
\nonumber \\
&\qquad +  2 p^{(\alpha)}_{22} [ \boldsymbol{\hat{x}}\otimes\boldsymbol{\hat{x}} - \boldsymbol{\hat{y}}\otimes \boldsymbol{\hat{y}} +\mathsf{i}(\boldsymbol{\hat{x}}\otimes\boldsymbol{\hat{y}}+\boldsymbol{\hat{y}}\otimes\boldsymbol{\hat{x}}) ] \Big].
\end{align}
\end{subequations}

The no-slip boundary conditions on both particle surfaces, which ensure that the fluid velocity matches the particle velocity at the surface, are given by: 
\begin{align}
    \left . \mathbf{u} \right |_{r_{\alpha}=a} = \mathbf{U}^{\alpha},
\end{align}
where $\mathbf{U}^{\alpha}$ denotes the particle velocity. Following Happel and Brenner~\cite{happel2012low}, we construct three scalar quantities from the boundary conditions and expand them in spherical harmonics:
\begin{subequations} 
\label{eqn:scalar_identities0}
\begin{align}
    \label{eqn:scalar_identities}
    \left . (\mathbf{u} \cdot \boldsymbol{\hat{r}}_{\alpha} ) \right |_{r_{\alpha}=a}  = \mathbf{U}_{\alpha} \cdot \boldsymbol{\hat{r}}_{\alpha} = \sum_{\substack{m=0 \\ n=m}}^{\infty} \chi^{(\alpha)}_{mn} Y^{m}_{n}(\theta_{\alpha},\phi), & \\
    \label{eqn:scalar_identities2} \left . \partial_{r_{\alpha}} ( \mathbf{u} \cdot \boldsymbol{\hat{r}}_{\alpha}) \right |_{r_{\alpha}=a}  = - \boldsymbol{\nabla} \cdot \mathbf{U}_{\alpha} = \frac{1}{a} \sum_{\substack{m=0 \\ n=m}}^{\infty} \psi^{(\alpha)}_{mn} Y^{m}_{n}(\theta_{\alpha},\phi), & \\
    \label{eqn:scalar_identities3} \left . \left ( \mathbf{r}_{\alpha} \cdot \boldsymbol{\nabla} \otimes \mathbf{u} \right ) \right |_{r_{\alpha}=a} = \mathbf{r}_{\alpha} \cdot \boldsymbol{\nabla} \otimes \mathbf{U}_{\alpha} = \sum_{\substack{m=0 \\ n=m}}^{\infty} \zeta^{(\alpha)}_{mn} Y^{m}_{n}(\theta_{\alpha},\phi). &
\end{align}
\end{subequations}
From \cref{eqn:scalar_identities0}, the expansion coefficients $p^{\alpha}_{mn}$, $q^{\alpha}_{mn}$, and $v^{\alpha}_{mn}$ can be directly related to the newly defined $\chi^{\alpha}_{mn}$, $\psi^{\alpha}_{mn}$, and $\zeta^{\alpha}_{mn}$.

Once the velocities on the surfaces of the spheres, $\mathbf{U}$, are specified, we can use \cref{eqn:forces_from_twinmultipole} to compute the forces, torques, and stresslets acting on each particle. From these results and the definition in \cref{eqn:res_tensors_def2}, it is possible to determine the resistance functions that make up the tensors $\mathbf{A}$, $\mathbf{B}$, $\mathbf{C}$, $\mathbf{G}$, $\mathbf{H}$, and $\mathbf{M}$ in \cref{eqn:res_tensors}. Calculating these 20 resistance functions is a complex and intricate process. The approach involves solving simpler, isolated problems with different prescribed velocities on the particle surfaces, which allow the individual components to be determined systematically. In Appendix~\ref{sec:appen_resistance}, we provide a detailed example of the calculation for the resistance functions associated with the longitudinal mode of the linear force-velocity coupling. For a comprehensive derivation of all resistance functions, readers are referred to the seminal works of Jeffrey and Onishi~\cite{jeffrey1984calculation,jeffrey1992calculation} and the corrections documented by Townsend~\cite{townsend2023generating}.

%

%It is important to note that the resulting two-body grand resistance is an approximation, as the scalar functions are derived from a series expansion. Nonetheless, the singular terms that arise when the surface-to-surface distance approaches zero are explicitly extracted and expressed in analytical form. Consequently, each resistance function comprises two components: a singular term that diverges as particles come into contact and a non-singular term represented by a rapidly converging series expansion.

Once all resistance functions are determined, the two-body grand resistance matrix can be used to assemble the total near-field grand resistance matrix, $\mathcal{R}^{\text{nf}}$, in a pairwise manner. Each sub-block of $\mathcal{R}^{\text{nf}}$ is constructed as follows
\begin{align}
{\mathbf{R}^{\text{nf}}_{\mathcal{F}\mathcal{U}}}
= 
\begin{bmatrix}
\displaystyle{\sum_{k \neq 1}^N}  \mathbf{R}^{11,k}_{\mathcal{F}\mathcal{U}} & \mathbf{R}^{12}_{\mathcal{F}\mathcal{U}} & \dots & \mathbf{R}^{1N}_{\mathcal{F}\mathcal{U}} \\
\mathbf{R}^{21}_{\mathcal{F}\mathcal{U}} & \displaystyle{\sum_{k \neq 2}^N} \mathbf{R}^{22,k}_{\mathcal{F}\mathcal{U}}  &  & \mathbf{R}^{2N}_{\mathcal{F}\mathcal{U}} \\
\vdots & & \ddots & \vdots\\
\mathbf{R}^{N1}_{\mathcal{F}\mathcal{U}} & \mathbf{R}^{N2}_{\mathcal{F}\mathcal{U}} & \dots & \displaystyle{\sum_{k \neq N}^N}  \mathbf{R}^{NN,k}_{\mathcal{F}\mathcal{U}} 
\end{bmatrix}.
\end{align}
Here, the off-diagonal blocks $\mathbf{R}^{ij}_{\mathcal{F}\mathcal{U}}$ correspond to the off-diagonal blocks of the two-body resistance matrix $\mathbf{R}^{2\text{-body}}_{\mathcal{F}\mathcal{U}}$ computed for the particle pair $(i,j)$. That is, each off-diagonal block in the total resistance matrix is extracted from a two-body interaction between the respective particles. The dimension of each block in the full $\mathbf{R}_{\mathcal{F}\mathcal{U}}$ matrix is thus half the dimension of a single two-body resistance. Similarly, the diagonal blocks $\mathbf{R}^{ii,k}_{\mathcal{F}\mathcal{U}}$ are constructed from the diagonal sub-block of $\mathbf{R}^{2\text{-body}}_{\mathcal{F}\mathcal{U}}$ computed for the pair $(i,k)$ , summing over all $k \neq i$. This accounts for the indirect contribution of all surrounding particles to the drag force experienced by particle $i$. Notably, since each two-body resistance is symmetric under the exchange of the two interacting particles, it does not matter which sub-block is extracted when assembling the full system-wide matrix. The other sub-blocks, such as $\mathbf{R}^{\text{nf}}_{\mathcal{S}\mathcal{U}}$, $\mathbf{R}^{\text{nf}}_{\mathcal{F}\mathcal{E}}$, and $\mathbf{R}^{\text{nf}}_{\mathcal{S}\mathcal{E}}$, are assembled in a similar manner.

%%%%%%%%
\section{Particle Dynamics}
%%%%%%%%

In this section, we use the hydrodynamic coupling described in the previous section to put together the equations of motion governing the colloidal particles.

%%%%%%%%
\subsection{Overdamped Motion}
%%%%%%%%

Assuming viscous forces dominate inertial effects, the particles undergo overdamped motion. The particles dynamics are described by the force-balance equation:
\begin{align}
    \label{eqn:eom}
    0 &= \mathcal{F}^P + \mathcal{F}^{H} + \mathcal{F}^B.
\end{align}
where $\mathcal{F^P}$, $\mathcal{F}^H$, and $\mathcal{F}^B$ are generalized force vectors representing particle-particle interactions, hydrodynamic drag, and Brownian (thermal) forces, respectively. Using the grand resistance matrix definition~\eqref{eqn:granres_def}, the hydrodynamic drag force is given by:
\begin{align}
        \label{resmatrix}
        \mathcal{F}^{H} &= - \mathbf{R}_{\mathcal{FU}} \cdot (\mathcal{U}-\mathcal{U}^{\infty}) + \mathbf{R}_{\mathcal{FE}} \cdot \mathcal{E}^{\infty}.  
\end{align}
The Brownian forces are instantaneously correlated through the HIs, with their variance determined by the fluctuation-dissipation theorem~\cite{kubo1966, Deutch1971fluct-diss}.

When considering a discrete-time simulation with time-step size $\Delta t$, this correlation can be expressed as~\cite{Ermak1978, Bossis1987}
\begin{align}
        \langle \mathcal{F}^B(t_i) \mathcal{F}^B(t_j) \rangle - \langle \mathcal{F^B}(t_i)\rangle  \langle \mathcal{F}^B(t_j)\rangle = \frac{2 k_{\mathrm{B}} T}{\Delta t} \mathbf{R}_{\mathcal{FU}} \delta_{ij},
\end{align}
where $\delta_{ij}$ indicates that the noise is uncorrelated at different time steps. The indices $i$ and $j$ label discrete time steps, and the angled brackets denote a time average.

To compute the Brownian contributions, we must specify by which convention we evaluate them~\cite{van1981ito}. For numerical studies, a common choice is the It{\^o} convention. At each (discrete) time $t_i$, we compute $\mathbf{R}_{\text{FU}}$ using the relative particle positions $\mathbf{r}(t_i)$. Following this convention, the Brownian forces acquire a non-zero average
\begin{align}
        \langle \mathcal{F^B}(t_i) \rangle &= k_{\mathrm{B}} T~\mathbf{R}_{\mathcal{FU}} \cdot \bm{\nabla}_{\mathbf{X}} \mathbf{R}^{-1}_{\mathcal{FU}} ,
\end{align}
which is usually called ``Brownian drift''. Here, $\bm{\nabla}_{\mathbf{X}}$ denotes the gradient with respect to all particle positions. Physically, its presence is required to ensure stationarity under the Gibbs–Boltzmann distribution and to generate particle configurations with the correct statistics at equilibrium~\cite{fiore2019fast}.

Combining these terms, the particle displacement equation becomes:
\begin{align} \label{eq:particle_disp}
    \mathcal{U} - \mathcal{U}^{\infty} &=  \mathbf{R}_{\mathcal{FU}}^{-1} \cdot (\mathcal{F^P} + \mathbf{R}_{\mathcal{FE}} \cdot \mathcal{E}^{\infty}) \nonumber \\
    &+ \sqrt{\frac{2k_{\mathrm{B}} T}{\Delta t}} \mathbf{R}_{\mathcal{FU}}^{-1/2} \cdot \mathcal{\psi} + k_{\mathrm{B}} T \, \nabla \cdot \mathbf{R}_{\mathcal{FU}}^{-1},
\end{align}
where $\mathcal{\psi}$ is a vector of random variables with zero mean and unit variance. This equation holds exactly when the full grand resistance matrix $\mathcal{R}$ is known. However, in practice, $\mathcal{R}$ is generally not accessible in its entirety. Instead, we construct approximations of $\mathcal{R}$ using the formalism developed earlier, extracting the relevant sub-blocks as needed. Alternatively, using the mobility formalism, the resistance blocks can be expressed as:
\begin{align} \label{eq:mobility-resistance-blocks}
    \mathbf{R}_{\mathcal{FU}}^{-1} &= \mathbf{M}_{\mathcal{UF}} - \mathbf{M}_{\mathcal{US}} \ \mathbf{M}_{\mathcal{ES}}^{-1} \ \mathbf{M}_{\mathcal{EF}} \\ 
    \mathbf{R}_{\mathcal{FE}}^{-1} &= \mathbf{M}_{\mathcal{EF}} - \mathbf{M}_{\mathcal{ES}} \ \mathbf{M}_{\mathcal{US}}^{-1} \ \mathbf{M}_{\mathcal{UF}}.
\end{align}

%%%%%%%%
\subsection{Hydrodynamic Model}
%%%%%%%%

In Stokesian Dynamics~\cite{brady1988stokesian,sierou2001accelerated,banchio2003accelerated,wang2016spectral,fiore2019fast}, HIs are modeled using both mobility and resistance formalisms, as outlined in Section~\ref{sec:II}. The method aims to balance computational efficiency with accuracy by combining a far-field approximation, which captures long-range interactions, with a near-field correction, which accounts for short-range hydrodynamics, both of which are analytic.

The far-field mobility matrix, $\mathcal{M}^{\text{ff}}$, is constructed using a truncated multipole expansion. While this provides an efficient way to approximate many-body interactions, it lacks accuracy at small separations. This is particularly the case for separations that place the particles in the lubricating flow regime, where hydrodynamic forces diverge. To address this, SD introduces a near-field resistance correction, $\mathcal{R}^{\text{nf}}$, derived from detailed two-body solutions. This ensures that even at close particle distances, HIs remain well-resolved. This hybrid approach provides a computationally feasible framework while retaining essential near-field physics, thus avoiding the need to explicitly resolve the full many-body problem.

The complete grand resistance matrix, $\mathcal{R}$, is then assembled as~\cite{brady1988stokesian}:
\begin{align} \label{eq:total_grandres_SD}
    \mathcal{R} = (\mathcal{M}^{\text{ff}})^{-1} + \mathcal{R}^{\text{nf}} - \mathcal{R}^{\text{ff}}_{\text{2B}}.
\end{align}
Here, $(\mathcal{M}^{\text{ff}})^{-1}$ provides a first approximation based on the far-field mobility, while $\mathcal{R}^{\text{nf}}$ introduces the necessary short-range corrections. To avoid double-counting, the two-body far-field resistance, $\mathcal{R}^{\text{ff}}_{\text{2B}}$, is subtracted~\cite{durlofsky1987dynamic}.

Although this formulation successfully bridges long- and short-range HIs, it has its limitations. In particular, the pairwise resistance corrections in $\mathcal{R}^{\text{nf}}$ neglect higher-order many-body effects beyond two-particle interactions. Similarly, while $\mathcal{M}^{\text{ff}}$ captures leading-order long-range interactions, truncation errors may still persist in dense suspensions. These constraints will be examined in detail in Section~\ref{sec:limitations}, where we assess the trade-offs inherent to this approximation.

%%%%%%%%
\section{Results and Discussion} \label{sec:result}
%%%%%%%%

In this section, we discuss the performance, accuracy, and scalability of Stokesian Dynamics compared to the Rotne-Prager-Yamakawa~\cite{rotne1969variational, swan2016rapid, fiore2017rapid} approximation and other advanced hydrodynamic techniques~\cite{wilson2013stokes}. Additionally, we discuss the limitations of SD and explore modern approaches for enhancing its accuracy and efficiency.

%%%%%%%%
\subsection{Comparative Performance: RPY vs. Stokesian Dynamics}
%%%%%%%%

\begin{figure}[!htb]
\centering
\includegraphics[width=85mm]{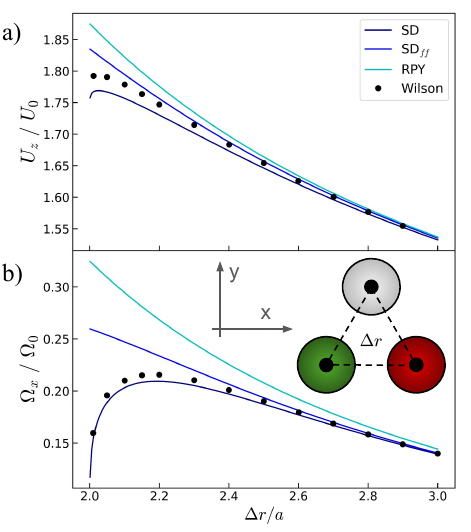}
\caption{\label{fig:3body_1}Comparison of Rotne-Prager-Yamakawa (RPY), Stokesian Dynamics (SD) without lubrication (SD$_\text{ff}$), full SD, and Wilson’s three-body results~\cite{wilson2013stokes}. We show the effect of HIs between three spherical particles (radius $a$) arranged in an equilateral triangle in the $xy$ plane, when these are subjected to identical out-of-plane forces (in the $z$-direction), see inset. (a) The translational velocity $U_z$ and (b) the angular velocity $\Omega_x$, which are the same for all spheres by symmetry. Both quantities are plotted as a function of the reduced interparticle distance $\Delta r / a$. The translational velocity $U_z$ is normalized by the velocity of an isolated sphere $U_0 = \Vert \mathbf{F} \Vert /(6\pi\eta a)$, while the angular velocity $\Omega_x$ is normalized by $\Omega_0 = \Vert \mathbf{F} \Vert/(6\pi\eta a^2)$, with $\mathbf{F}$ the applied force and appropriate scaling in $a$.}
\end{figure}

\begin{figure}[!htb]
\centering
\includegraphics[width=85mm]{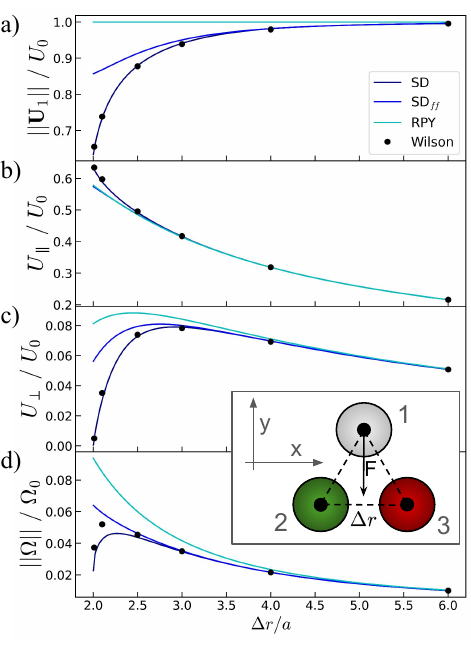}
\caption{\label{fig:3body_2}The same geometrical setup as in Fig.~\ref{fig:3body_1} is considered to compare hydrodynamic approximations. In this case, a force is applied to a single sphere (\#1) within the plane of the triangular arrangement, directed radially inward toward the centroid of the equilateral triangle, see the inset. The other two spheres (\#2 and \#3) remain force-free,~\textit{i.e.}, their motion is entirely induced by HIs. Quantities and axis normalizations are as in Fig.~\ref{fig:3body_1}. Panel (a) shows the ($y$-component) translational velocity $U_1$ of the driven sphere, while panels (b) and (c) show the induced translational velocities of the neighboring force-free spheres, resolved into components parallel ($U_{\parallel}$) and normal ($U_{\perp}$) to the applied force, respectively. Panel (d) presents the magnitude of the induced angular velocities $\Vert \boldsymbol{\Omega} \Vert$ of the force-free neighboring particles.}
\end{figure}

We investigate the accuracy of SD relative to the computationally cheaper RPY approximation by analyzing two distinct three-body configurations in Stokes flow, following work by Wilson~\cite{wilson2013stokes}. In the first setup, three spherical particles are arranged in an equilateral triangle within the $xy$ plane, and all particles are forced out of plane by applying a unit normal force in the $z$-direction. The resulting translational $U_z$ and angular velocities $\Omega_x$ are plotted in Figures~\ref{fig:3body_1}a and~\ref{fig:3body_1}b as functions of the interparticle distance $\Delta r$. In the second setup, the same equilateral triangle configuration is studied, but here an in-plane force is applied to a single particle, pushing it toward the center of the triangle. The velocity $\mathbf{U}_1$ of the forced particle is shown in Figure~\ref{fig:3body_2}a, while the induced velocities on the other two spheres --- resolved into parallel ($U_{\parallel}$) and normal ($U_{\perp}$) components with respect to the applied force --- are plotted in Figures~\ref{fig:3body_2}c and~\ref{fig:3body_2}d, respectively. Figure~\ref{fig:3body_2}d presents the magnitude of the induced angular velocities $\Vert \boldsymbol{{\Omega}} \Vert$ of the force-free particles. In each numerical experiment, we compare results from RPY, SD without lubrication corrections (SD$_\text{ff}$), full SD, and Wilson’s three-body method~\cite{wilson2013stokes} --- considered exact for these setups.

As expected, full SD, which includes lubrication corrections, exhibits the closest agreement with Wilson’s results across all cases, not only in magnitude but also in capturing the overall trends. This highlights the importance of lubrication effects for capturing near-field HIs, particularly as particle separations decrease. In contrast, the RPY approximation provides good accuracy only at sufficiently large interparticle distances. The threshold distance beyond which RPY begins to fail qualitatively depends on the specific problem being studied. For example, in the first setup, RPY performs reasonably well for translational velocities but exhibits more significant departures for angular velocities, as seen in \cref{fig:3body_1}b. This discrepancy arises because angular velocities are more sensitive to higher-order HIs, which the RPY approximation does not take into account.

A more notable limitation of RPY emerges in the second problem, where the method fails to capture the influence of surrounding particles on the self-mobility of the forced sphere. As shown in Figure~\ref{fig:3body_2}a, the RPY self-term remains constant at unity, lacking any contribution from the nearby particles. In contrast, SD correctly accounts for many-body hydrodynamic coupling, resulting in deviations from unity and a more realistic prediction of the particle's velocity. Minor deviations between SD and Wilson's reference result are also observed in the induced forces and angular velocities of the neighboring spheres (\ref{fig:3body_2}b–d). However, they are always much smaller than those of RPY, further underscoring the limitations of the pairwise approximation.

In addition to accuracy, computational cost is an important consideration when selecting a method. The RPY approximation scales linearly with the number of particles $N$, making it computationally efficient even for very large systems. In contrast, traditional implementations of SD require the inversion of dense matrices, leading to a cost that scales as $O(N^3)$ for direct methods. While SD is significantly more expensive, its ability to resolve near-field and many-body hydrodynamic effects justifies its use when high accuracy is required.

The choice between RPY and SD ultimately depends on the system under investigation and the required level of precision. Ideally, both methods should be available for comparative testing to assess whether the RPY approximation is sufficient for the phenomenon of interest~\cite{varga2018normal, torre2023hydrodynamic}, which is facilitated by our implementation~\cite{torre2025pythonjaxbasedfaststokesiandynamics}, offering both approaches in a unified framework. In many cases, RPY may be adequate at large separations or when angular velocities and near-field effects are not critical~\cite{varga2018normal}. However, when precise hydrodynamic coupling is essential, as demonstrated here, SD remains the method of choice.

%%%%%%%%
\subsection{Addressing Limitations in Conventional Stokesian Dynamics} \label{sec:limitations}
%%%%%%%%

Before turning to efficiency in Section~\ref{sec:IVc}, we will comment on the accuracy limitations of Stokesian Dynamics. These arise under special conditions where the assumptions or approximations within SD fail to adequately represent the underlying physics. By understanding these limitations, we can identify potential strategies for improving the method.

A common source of inaccuracy in SD simulations, particularly when comparing to experimental results, arises from the necessity of using periodic boundary conditions (PBCs) to simulate bulk suspensions. PBCs are an essential tool in computational studies, allowing for the efficient modeling of macroscopic properties while avoiding the need for unrealistically large simulation domains. However, they also introduce artificial long-range HIs between the periodic images of particles~\cite{kim2013microhydrodynamics}. These effects are particularly pronounced in dilute systems, where periodicity can lead to systematic deviations from the true bulk behavior~\cite{phillips1988hydrodynamic}. While necessary for computational feasibility, these artifacts can be mitigated by applying finite-size corrections or accounting for microstructural effects, as shown in structure-factor-based corrections for sedimentation dynamics~\cite{sangani1994method}.

Separately, SD relies on a truncated multipole expansion to capture HIs. This expansion provides excellent accuracy for well-separated particles and is further complemented by lubrication corrections at very short distances. However, an intermediate regime exists where neither low-order multipoles nor lubrication effects dominate, leading to potential inaccuracies~\cite{phillips1988hydrodynamic}. This regime is particularly relevant in moderately dense suspensions, where particle separations are small enough for near-field interactions to matter but not sufficiently close for lubrication corrections to fully resolve the hydrodynamics. Addressing this limitation would require extending the multipole expansion to higher orders, which would come at the cost of increased computational complexity.

A clear example illustrating both periodicity effects and intermediate-range inaccuracies is the sedimentation velocity of a suspension of hard spheres with a fluid-like spatial organization. At low volume fractions $\phi$, the sedimentation velocity calculated using SD shows a characteristic $N^{-1/3}$ dependence on the number of particles~\cite{phillips1988hydrodynamic,sierou2001accelerated}, reflecting the influence of PBCs. This scaling, caused by the long-range effects of periodic images, can distort the dynamics compared to those of a truly random, bulk suspension. Mo and Sangani~\cite{sangani1994method} demonstrated that this periodicity-induced discrepancy can be corrected by considering the structure factor, which accounts for the suspension's microstructure, and applying an appropriate correction term.

At higher $\phi$, the sedimentation dynamics of hard spheres present a different challenge. In this regime, the restricted fluid flow between tightly packed particles reduces the influence of lubrication interactions, while mid-range HIs become increasingly important. As a result, the accuracy of SD depends on the inclusion of higher-order hydrodynamic moments in the multipole expansion~\cite{phillips1988hydrodynamic, sierou2001accelerated}. For instance, achieving even $10\%$ accuracy at $\phi=0.5$ requires including up to hexadecapole term~\cite{phillips1988hydrodynamic}. However, the computational cost of including more moments grows rapidly, as it increases both the number of unknowns and the complexity of the calculations~\cite{durlofsky1987dynamic}. To make these enhancements practical, faster implementations of SD are essential, which make the simulation of many particles feasible, as well as pave the way for the development of increasingly accurate approaches to hydrodynamic simulations.

%%%%%%%%
\subsection{Recent Advancements that enable Faster and Scalable Stokesian Dynamics}\label{sec:IVc}
%%%%%%%%

Taking on the challenges identified in the previous subsection, the field has focused improving the efficiency and scalability of SD~\cite{sierou2001accelerated, banchio2003accelerated, wang2016spectral, fiore2019fast, MA2022115496}. SD's computational expense has historically limited its application to large systems and dense suspensions. In this section, we discuss several advancements that helped remedy this situation. We also propose a novel hybrid framework that combines SD methods with neural network-based approaches.

One major breakthrough came with Accelerated Stokesian Dynamics (ASD), introduced by Sierou and Brady~\cite{sierou2001accelerated}. This approach utilized iterative solvers and Ewald summation to reduce computational scaling from $O(N^3)$ to $O(N\log{N})$. ASD made it feasible to simulate larger particle systems while retaining the essential features of the SD framework. Later, Banchio and Brady extended ASD to include Brownian motion~\cite{banchio2003accelerated}, ensuring that thermal fluctuations and equilibrium properties were faithfully captured.

More recently, Fiore and Swan~\cite{fiore2019fast} developed Fast Stokesian Dynamics, which optimized performance further by employing spectral Ewald methods and matrix-free operations, achieving $O(N)$ scaling. They also made their algorithm widely available through integration in the \textbf{\texttt{HOOMD}}~\cite{ANDERSON2020109363} simulation package. FSD is particularly effective for large-scale simulations, allowing accurate modeling of HIs in systems with thousands of particles while leveraging GPU acceleration for efficient parallel computations. Our implementation~\cite{torre2025pythonjaxbasedfaststokesiandynamics} extends this approach by providing a fully modular Python framework with support for both open and periodic boundary conditions, as well as easily implementable interaction potentials.

Neural network-based methods have recently emerged as promising tools for simulating HIs in particle suspensions. Ma~\textit{et al}.~\cite{MA2022115496} introduced the Hydrodynamic Interaction Graph Neural Network (HIGNN), a framework designed to predict particle dynamics in suspensions. By representing particles as nodes in a graph and HIs as edges, HIGNN efficiently models the mobility tensor with high accuracy. The network incorporates higher-order connectivity to account for many-body effects, capturing both long-range and short-range HIs. HIGNN's key advantage is its scalability --- it can simulate suspensions with tens of thousands of particles on modest computational hardware, while maintaining an accuracy that is comparable to traditional SD methods.

Inspired by this development and wanting to circumvent SD's current limitations, we propose a hybrid framework that uses SD to train a neural network for predicting many-body hydrodynamic corrections. The key idea is to decompose the mobility matrix into analytically known two-body terms and an unknown many-body contribution, which the neural network learns.

In the absence of shear and thermal motion, the translational velocity of each particle $\alpha$ can be expressed in terms of the mobility tensor as:
\begin{align} 
\mathbf{U}_{\alpha} = \sum_{\beta} \boldsymbol{\mu}_{\alpha \beta} \cdot \mathbf{F}_{\beta}, 
\end{align}
where $\boldsymbol{\mu}_{\alpha \alpha}$ represents the self-mobility of particle $\alpha$, and $\boldsymbol{\mu}_{\alpha \beta}$ accounts for HIs between particles $\alpha$ and $\beta$. Using SD, we can obtain these mobilities by extracting the relevant sub-blocks of $\mathbf{R}_{\mathcal{FU}}^{-1}$. However, we can also construct an approximate model using only two-body results, where the pair mobilities are taken from exact two-body solutions~\cite{jeffrey1984calculation} (including lubrication), and the self-mobility is approximated by the identity matrix:
\begin{align}\label{eq:rpy_lub}
\mathbf{U}_{\alpha} \approx   \mathbf{I}  \cdot \mathbf{F}_{\alpha} + \sum_{\beta \neq \alpha} \boldsymbol{\mu}^{\text{pair-2b}}_{\alpha \beta} \cdot \mathbf{F}_{\beta}.
\end{align}
This approximation, which we denote as RPY$_{\text{lub}}$, partially incorporates lubrication and avoids matrix inversion, but entirely neglects many-body interactions.

To recover many-body effects, we introduce a correction matrix $\mathbf{G}_{\alpha \beta}$, defined such that
\begin{align}
\boldsymbol{\mu}_{\alpha \beta} = \boldsymbol{\mu}^{\text{pair-2b}}_{\alpha \beta} \cdot \mathbf{G}_{\alpha \beta},
\end{align}
where $\mathbf{G}_{\alpha \beta}$ encodes many-body interactions. This matrix is expected to be close to the identity when other particles are far from the interacting pair $(\alpha,\beta)$, thus capturing how many-body effects modify two-body interactions. Instead of learning the full mobility matrix, we train a neural network to predict $\boldsymbol{\mu}_{\alpha \alpha}$ and $\mathbf{G}_{\alpha \beta}$ (see Appendix~\ref{sec:appen_nn}), while the rest of the hydrodynamic contributions remain analytically known as in the RPY$_{\text{lub}}$ approximation. The identity approximation in~\cref{eq:rpy_lub} is replaced by the learned self-mobility:
\begin{align}
\mathbf{U}_{\alpha} \approx \boldsymbol{\mu}_{\alpha \alpha} \cdot \mathbf{F}_{\alpha} + \sum_{\beta \neq \alpha} \boldsymbol{\mu}^{\text{pair-2b}}_{\alpha \beta} \cdot \mathbf{G}_{\alpha \beta} \cdot \mathbf{F}_{\beta},
\end{align}
which ensures that both self- and many-body corrections are incorporated.

\begin{figure}[!htb] \centering \includegraphics[width=85mm]{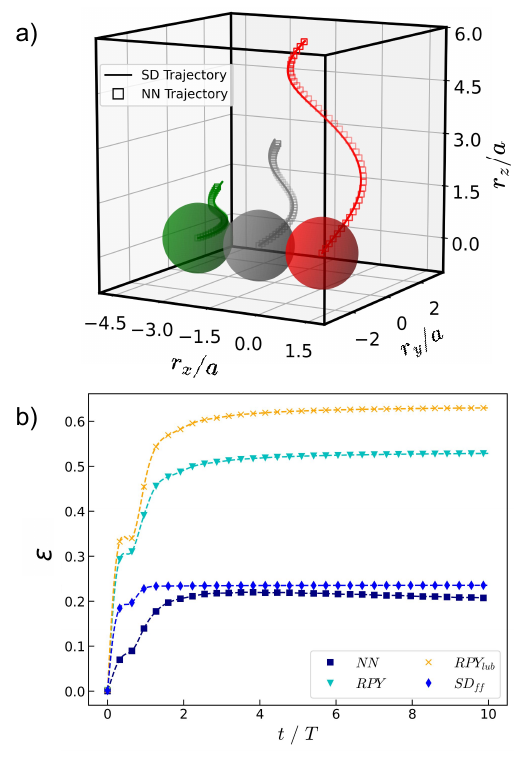} 
\caption{\label{fig:nn_trajectories}Proof of principle for the quality of our neural-network hydrodynamic solver. The system consists of three particles initially on a line, one of which (red) is subjected to an applied (reduced) force of the form $\mathbf{F}=(\cos{t},\sin{t}, 1)$ causing it to move. The other two (green and gray) move due to HIs alone. (a) Trajectories during the first period of oscillation, comparing SD (solid lines) and NN (open squares). Coordinates are expressed in units of particle radius $a$. (b) The average error per particle $\epsilon$ for the same system, as a function of reduced time $t/T$, where $T=2 \pi$. The error is defined as the norm of the displacement difference between the method of interest and SD, $\epsilon = (3a)^{-1} \sum^{3}_{\alpha=1} \Vert \mathbf{r}_{\text{method}}^\alpha - \mathbf{r}_{\text{SD}}^\alpha \Vert$, averaged over the three particles; indexed using $\alpha$. Squares represent errors from our neural network-corrected model (NN), triangles correspond to standard RPY, crosses indicate RPY with lubrication (RPY$_{\text{lub}}$), and diamonds denote SD without lubrication corrections (SD$_{\text{ff}}$). Simulations were performed with a timestep $\Delta t=0.01$ over 10 periods $T$ of oscillation.}
\end{figure}

As a proof of concept, we apply this method to a simple system of three spheres initially aligned along the $x$-axis in an unbounded fluid. The spheres are spaced equidistantly $2.005 \times a$ apart (center-to-center distance). Only one of the particles is subjected to a time-dependent force $\mathbf{F}=(\cos{t},\sin{t}, 1)$, while the other two are force-free and move solely due to HIs. To evaluate the accuracy of the approach, we compare the resulting particle trajectories obtained with SD, the neural network-corrected model, the RPY approximation with and without lubrication corrections, and SD without lubrication corrections. The results, shown in Figure~\ref{fig:nn_trajectories}, demonstrate that the neural network-corrected model yields the smallest trajectory errors, closely matching SD across all timescales.

At short times, the NN-corrected model benefits from its ability to refine lubrication effects and many-body HIs, significantly reducing errors. At longer times, a notable trend emerges: while the NN-corrected model maintains the lowest error, it gradually approaches the error of SD$_{\text{ff}}$. This is consistent with the fact that as particles separate over time, lubrication forces become negligible, and only many-body HIs remain relevant. Since SD without lubrication still captures these many-body interactions, its long-time error trends align closely with those of the NN model. Interestingly, the standard RPY approximation produces smaller errors than RPY$_{\text{lub}}$, likely due to an overestimation of HIs when neglecting many-body effects~\cite{durlofsky1987dynamic}; lubrication corrections amplify these errors instead of mitigating them.

By combining exact two-body results with machine-learned many-body corrections, this approach offers a computationally efficient alternative to SD while retaining good accuracy. The ability to bypass expensive matrix inversions while preserving the essential many-body physics makes this framework particularly promising for large-scale simulations of particle suspensions. We have included details on the implementation and training of our neural network in Appendix~\ref{sec:appen_nn}

It should be noted that, in this study, we did not focus on optimizing the selection of training data for our neural network model. It is well-established that the quality and diversity of training data significantly influence a model's accuracy~\cite{fan2022effects,budach2022effects}. Therefore, by strategically selecting representative and meaningful data, the model's ability to capture many-body HIs can be enhanced, leading to improved predictive performance. Exploring this potential for improvement in a large-scale system is left for future work.

Additionally, in dilute dispersions, two-body HIs provide a sufficiently accurate description of particle motion~\cite{rotne1969variational, happel2012low, kim2013microhydrodynamics}. As the particle density increases, many-body effects become significant and must be accounted for~\cite{happel2012low,kim2013microhydrodynamics}. While these interactions are theoretically long-ranged, their practical influence is limited to a finite region around each particle. This is a consequence of the way flow disturbances propagate and decay in a suspension. The velocity field induced by one particle extends through the fluid, but successive interactions with other particles redistribute and dissipate momentum, causing the net effect of distant particles to become less relevant in practice~\cite{durlofsky1987dynamic}. This physical behavior allows us to introduce a hard cutoff in dynamical simulations: beyond a certain distance, particles interact only through their analytically known two-body kernel, while many-body corrections are applied within a localized neighborhood. This approach significantly reduces computational complexity, as the dominant many-body effects are effectively captured within this finite environment.

%%%%%%%%
\section{Conclusion}
%%%%%%%%

In summary, Stokesian Dynamics~\cite{brady1988stokesian, sierou2001accelerated, banchio2003accelerated, wang2016spectral, fiore2019fast,torre2025pythonjaxbasedfaststokesiandynamics} has proven itself as a cornerstone method for studying hydrodynamic interactions in particle suspensions. This review has traced the method's evolution, highlighting its successes, limitations, and the modern advancements aimed at overcoming the identified challenges.

Our comparative analysis demonstrates that SD consistently outperforms simpler approximations in terms of accuracy, such as the Rotne-Prager-Yamakawa tensor~\cite{rotne1969variational, swan2016rapid, fiore2017rapid}. This holds especially for scenarios where near-field interactions and many-body hydrodynamic couplings are critical~\cite{gelhydroJoost,torre2023hydrodynamic}. However, the limitations of conventional SD become apparent in specific contexts, such as sedimentation dynamics of random suspensions at both low and high volume fractions~\cite{phillips1988hydrodynamic, sangani1994method, sierou2001accelerated}. At low concentrations, the artificial periodicity introduced by commonly applied boundary conditions distorts sedimentation velocities~\cite{sangani1994method}. In contrast, at higher concentrations, the absence of higher-order multipole terms reduces the method’s accuracy in systems where lubrication effects are less significant~\cite{phillips1988hydrodynamic}. We have indicated that these challenges underscore the need for methods that address intermediate-range interactions and reduce the computational cost of higher-order expansions.

Modern advancements, such as Accelerated and Fast Stokesian Dynamics~\cite{sierou2001accelerated,banchio2003accelerated,wang2016spectral,fiore2019fast}, have revitalized the use of SD by improving its base algorithmic scaling. Presently, simulation of large systems with linear or near-linear scaling in the particle number are possible. This makes SD suited for the study of dense suspensions and systems with features on many scales,~\textit{e.g.}, colloidal gels~\cite{torre2023hydrodynamic}.

The integration of machine learning into the modeling of particle suspensions, exemplified by neural network frameworks like HIGNN~\cite{MA2022115496}, offers a new frontier for SD. We recognize that these approaches provide a pathway to bypass traditional bottlenecks, by leveraging data-driven models to predict interactions with high accuracy, while maintaining computational efficiency. Here, we provide a first look into the use of machine learning for SD. We propose a hybrid framework, combining the precision of SD with the efficiency of neural networks. Our approach exploits the strengths of both methodologies: SD generates efficiently high-quality training data for neural networks, which in turn provide corrections to analytical two-body interactions at the RPY level.

In conclusion, the trajectory of Stokesian Dynamics, from its original formulation~\cite{durlofsky1987dynamic,brady1988stokesian} to modern advancements~\cite{fiore2019fast,MA2022115496}, reflects the evolving challenges and demands of hydrodynamic simulations. By combining foundational methods with innovative computational techniques, the future of SD is looking bright~\cite{bolintineanu2014particle}.

%%%%%%%
\section*{Acknowledgements}
%%%%%%%

The authors acknowledge funding from NWO (grant OCENW.KLEIN.354) and IFPRI (grant CRR-118-01). We are grateful to Luca Leone, Gwynn Elfring, and John F. Brady for useful discussion. An open data package containing the means to reproduce the results of the simulations is available at: [DOI]

%%%%%%%
\section*{Author Contributions}
%%%%%%%

Author contributions: Conceptualization, J.d.G. \& K.W.T.; Methodology, K.W.T.; Numerical calculations, K.W.T., Validation, K.W.T.; Investigation,
K.W.T.; Writing --- Original Draft, K.W.T.; Writing --- Review \& Editing, J.d.G.; Funding Acquisition, J.d.G.;
Resources, J.d.G.; Supervision, J.d.G.

%%%
\bibliographystyle{aip}
\bibliography{reference}
%%%

\appendix

%%%%%%%
\section{\label{sec:appen_mobility}Elements of the grand Mobility }
%%%%%%%%

In this appendix we explicitly compute some of the matrix sub-blocks of the far-field grand mobility matrix $\mathcal{M}^{\text{nf}}$ defined in equation~\eqref{eqn:granmobility_def}. This will allow interested readers to gain a handle on the process and perform the remaining calculations themselves. Invoking linearity~\cite{durlofsky1987dynamic,kim2013microhydrodynamics}, we can generally write equation~\eqref{eqn:faxen1} as
\begin{align}
    \label{eqn:faxen1_explicit}
    U_i^{\alpha} = - \sum^{N}_{\beta=1} \left ( {a}^{\alpha \beta}_{ij} \ F_j^{\text{H},\beta} + \tilde{b}^{\alpha \beta}_{ij} \ T_j^{\text{H},\beta} + \tilde{g}^{\alpha \beta}_{ijk} \ S_{jk}^{\beta} \right ),
\end{align}
where we have assumed for simplicity that the background flow $\boldsymbol{u}^{\infty}$ is zero everywhere in the system. Here, $\boldsymbol{a}^{\alpha \beta}$, $\boldsymbol{\tilde{b}}^{\alpha \beta}$, and $\boldsymbol{\tilde{g}}^{\alpha \beta}$ are tensors, which components depend on the particles relative positions. To simplify the discussion, we focus only on the coupling between particles linear velocities and hydrodynamic drag forces, thus, ignoring contribution to the particles velocities coming from hydrodynamic drag torques and stresslets. In this reduced picture, the flow field generated by all particles, except particle $\alpha$, can be approximated as 
\begin{align}
    u'_i(\mathbf{r}) \approx - \frac{1}{8 \pi \eta} \sum^N_{\beta \ne \alpha} \left [ \left ( 1+\frac{a}{6} \nabla^2 \right ) J_{ij}(\mathbf{r}-\mathbf{r}^{\beta})  F_j^{\text{H},\beta} \right ].
\end{align}
Using this expression for the fluid flow in equation~\eqref{eqn:faxen1}, and comparing with equation~\eqref{eqn:faxen1_explicit}, we get the following expressions for the tensors $\boldsymbol{a}^{\alpha \beta}$:
\begin{align}
  \label{eqn:tensor_a}
    \gamma a^{\alpha \beta}_{ij} =
    \begin{cases} 
        \delta_{ij} & \text{if } \alpha = \beta, \\
        - (\frac{3}{4} + \frac{a^2}{4} \nabla^2 + \frac{a^4}{48} \nabla^4) J_{ij}(\mathbf{\Delta r}^{\alpha \beta}) & \text{if } \alpha \ne \beta,
    \end{cases}
\end{align}
where we have defined $\mathbf{\Delta r}^{\alpha \beta} \equiv \mathbf{r}^{\alpha} - \mathbf{r}^{\beta} $, and $\gamma = 6 \pi \eta a$.

In an infinite fluid, the reduced Green's function reads~\cite{durlofsky1987dynamic,kim2013microhydrodynamics}
\begin{align}
    \label{eqn:stokeslet}
     {J}_{ij}(\mathbf{\Delta r}^{\alpha \beta}) = \frac{1}{\Delta r} (\delta_{ij} + e_i e_j),
\end{align}
with $\Delta r$ the vector length, and $\mathbf{e} \equiv \mathbf{\Delta r}^{\alpha \beta} / \Delta r$. Exploiting the symmetries of the grand mobility matrix~\cite{jeffrey1984calculation,durlofsky1987dynamic,kim2013microhydrodynamics}, we can generally write the tensor $\boldsymbol{a}^{\alpha \beta}$ in terms of two scalars $y^a_{\alpha \beta}$ and $x^a_{\alpha \beta}$, mobility functions of the inter-particle distance:
\begin{align}
     \gamma a^{\alpha \beta}_{ij} = y^a_{\alpha \beta} \delta_{ij} + e_i e_j (x^a_{\alpha \beta} - y^a_{\alpha \beta}).
\end{align}
Next, we combine the above expression with~\eqref{eqn:tensor_a}, to obtain expressions for the scalar mobility functions,
\begin{align}
    x^a_{\alpha \beta}(\Delta r) &=
    \begin{cases} 
        1 & \text{if } \alpha = \beta, \\
        \frac{3}{2} {\Delta r}^{-1} - {\Delta r}^{-3} & \text{if } \alpha \ne \beta ;
    \end{cases}  \\
    y^a_{\alpha \beta}(\Delta r) &=
    \begin{cases} 
        1 & \text{if } \alpha = \beta, \\
        \frac{3}{4} {\Delta r}^{-1} + \frac{1}{2} {\Delta r}^{-3} & \text{if } \alpha \ne \beta.
    \end{cases}
\end{align}

By using the full equation~\eqref{eqn:velfield2}, instead of truncating it at the level of forces, we could extend the calculation and obtain the complete set of scalar mobility functions~\cite{kim2013microhydrodynamics}, and correspondent tensors. Once all tensors are obtained, the far-field grand-mobility matrix can be assembled and used to obtain the full set of particle generalized velocities and strain.

%%%%%%%
\section{\label{sec:appen_resistance}Elements of the grand Resistance Matrix }
%%%%%%%%

In this appendix, we compute the elements of one of the sub-block of the near-field grand resistance matrix $\mathcal{R}^{\text{nf}}$ defined in equation~\eqref{eqn:granmobility_def}. Similarly to the grand mobility matrix, the sub-blocks of the grand-resistance matrix in equations~\eqref{eqn:res_tensors} and~\eqref{eqn:res_tensors2} can be expressed in terms of scalar functions. For simplicity, we consider only the tensor $\mathbf{A}_{\alpha \beta}$. The latter can be expressed as
\begin{align}
     \gamma^{-1} A^{\alpha \beta}_{ij} = Y^A_{\alpha \beta} \delta_{ij} + e_i e_j (X^A_{\alpha \beta} - Y^A_{\alpha \beta}),
\end{align}
with $Y^A_{\alpha \beta}$, and $X^A_{\alpha \beta}$ the scalar resistance functions associated with the sub-block $\mathbf{A}_{\alpha \beta}$. Here, we can impose particles velocities such that the effect of each scalar resistances is isolated~\cite{jeffrey1984calculation}. We proceed by showing the approach to obtain the scalars $X^A_{1 1}$, and $X^A_{12}$. For a detailed calculation of the entire set of resistance functions we direct the interested reader to the original work by Jeffrey and Onishi~\cite{jeffrey1984calculation,jeffrey1992calculation}, and the subsequent revisions by Townsend~\cite{townsend2023generating} that corrected several small issues.

Without loss of generality, we can rotate the reference frame in order to have the vector connecting particle $\alpha$ with $\beta$ aligned with the $\mathbf{\hat{z}}$-axis. We then impose $\mathbf{U}_1 = - \mathbf{U}_2 = U \boldsymbol{\hat{z}}$, from which we can write equations~\eqref{eqn:scalar_identities},~\eqref{eqn:scalar_identities2}, and~\eqref{eqn:scalar_identities3} as
\begin{align}
     \chi^{\alpha}_{mn} &= U \delta_{m0} \delta_{n1} ; \\
     \psi^{\alpha}_{mn} &= \zeta^{\alpha}_{mn} = 0,
\end{align}
while the equation~\eqref{eqn:res_tensors_def2} for the force on particle $\alpha = 1$ reads:
\begin{align}
    \label{eq:lubrication_force_XA}
    \mathbf{F}_{1}^{\text{H}} = -\gamma U \left( X^A_{1 1} - X^A_{1 2} \right)  \boldsymbol{\hat{z}}.
\end{align}
Comparing the above result with equation~\eqref{eqn:forces_from_twinmultipole}, it is clear that the only quantities we need to compute the force are the pressure coefficients $p^{\alpha}_{0n}$. Following the steps of Jeffrey and Onishi~\cite{jeffrey1984calculation}[p.266], we combine equation~\eqref{eqn:lambsol2} with the scalar identities from equations~\eqref{eqn:scalar_identities0}, to obtain a pair of equations that allow us to have recursive relations for the unknown coefficients:
\begin{align}
    (n+1)(2n+1)v_{0n}^{\alpha} &= \frac{(n+1)}{2}p_{0n}^{\alpha} \nonumber \\
    & \quad - \frac{n \, t^{n+1}}{2n+3} \sum_{s=0}^\infty \binom{s+n}{n} t^s p^{3-\alpha}_{0s} ,
\end{align}
\begin{align}
    (n+2)U \delta_{1n} = \frac{n+1}{2n-1} p^{\alpha}_{0n} + \sum_{s=0}^\infty \binom{s+n}{n} t^{s+n-1} \Bigg[ \frac{n \, t^2}{2} p^{\alpha}_{0s} & \nonumber \\
    + \frac{n(2n+1)}{2(2n-1)} \frac{(n+s-2ns+2)}{(2s-1)(n+s)} p^{3-\alpha}_{0s} + n(2n+1) t^2 v^{3-\alpha}_{0s} \Bigg] . &
\end{align}
with the parameter $t = a / \Delta r$. By expanding the pressure and velocity coefficients as a double power series in $t$~\cite{jeffrey1984calculation}:  
\begin{align}
    p_{0n}^{\alpha} &= \frac{3}{2} U \sum^{\infty}_{p,q=0} P^{\alpha}_{npq} t^{p+q} \\
    v_{0n}^{\alpha} &= \frac{3}{4(2n+1)} U \sum^{\infty}_{p,q=0} V^{\alpha}_{npq} t^{p+q}  ,
\end{align}
we can further obtain the recursive relations
\begin{align}
    V_{npq} &= P_{npq} \nonumber \\
    & \quad - \frac{2n}{(n+1)(2n+3)} \sum^{q}_{s=1} \binom{s+n}{n} P_{s(q-s)(p-n-1)} ,
\end{align}
\begin{align}
    P_{npq} &= \sum^{q}_{s=1} \Bigg[ \frac{n(2n+1)(2ns-n-s+2)}{2(n+1)(2s-1)(n+s)} P_{s(q-s)(p-n+1)} \nonumber \\
    &\phantom{= \sum^{q}_{s=1} \Bigg[} - \frac{n(2n-1)}{2(n+1)} P_{s(q-s)(p-n-1)} \nonumber \\ 
    &\phantom{= \sum^{q}_{s=1} \Bigg[} - \frac{n(4n^2-1)}{2(n+1)(2s+1)} V_{s(q-s-2)(p-n+1)} \Bigg],
\end{align}
with $V_{n00} = P_{n00} = \delta_{1n}$. We have dropped the superscript $\alpha$, because the coefficient are the same for each particle~\cite{jeffrey1984calculation}. Once these are compute, we combine equation~\eqref{eq:lubrication_force_XA} with equation~\eqref{eqn:forces_from_twinmultipole} to obtain
\begin{align}
    \label{eq:XA_minus}
    \gamma \left ( X^A_{1 1} - X^A_{1 2} \right ) = \sum^{\infty}_{p,q=0} P_{1pq} t^{p+q}.
\end{align}
We now repeat the same calculation but imposing a different velocities for the pair of particles, namely $\mathbf{U}_1 = \mathbf{U}_2 = U \boldsymbol{\hat{z}}$, obtaining this time 
\begin{align}
     \chi^{\alpha}_{mn} &= (-1)^{3-\alpha} U \delta_{m0} \delta_{n1} ; \\
     \psi^{\alpha}_{mn} &= \zeta^{\alpha}_{mn} = 0.
\end{align}
It is possible to show~\cite{jeffrey1984calculation} that the coefficients solving this problem are the same as the ones before, multiplied by a factor of $(-1)^{n+p+q+\alpha}$, which implies that
\begin{align}
    \label{eq:XA_plus}
    \gamma \left ( X^A_{1 1} + X^A_{1 2} \right ) = \sum^{\infty}_{p,q=0} (-1)^{p+q} P_{1pq} t^{p+q}.
\end{align}
Combining the above result with equation~\eqref{eq:XA_plus}, it is straightforward to see that $X^A_{11}$ must a series only of even powers of $t$, while $X^A_{12}$ only odd powers. Thanks to this result, we can finally write both scalar resistance functions as
\begin{align}
    \gamma X^A_{1 1} &= \sum^{\infty}_{k=0} f_{2k} (2s)^{-2k} ;\\
    \gamma X^A_{1 2} &= -\sum^{\infty}_{k=0} f_{2k+1} (2s)^{-2k-1},
\end{align}
where we changed variable to $s=\Delta r / a $, and defined $f_k = 2^k \sum^{k}_{q=0} P_{1(k-q)q}$. These expression of the resistance functions are exact for any distance in principle, however, the convergence of each series become slower and slower as the particles gap get closer to contact.

Close to contact, a much better framework to compute the resistance functions is given by lubrication theory of Jeffrey~\cite{jeffrey1982low}, which gives the correct asymptotic scaling up to a constant value, and neglecting non-singular terms (which vanish at contact). To obtain explicit expression of this offsets and non-singular terms, Jeffrey and Onishi~\cite{jeffrey1984calculation} combined these asymptotic expressions, with their series expansions. For completeness, we report the final form of the resistance scalars close to contact, up to order $O(\xi \ln{ \xi^{-1} })$, and redirect the reader to the work of Jeffrey and Onishi~\cite{jeffrey1984calculation,jeffrey1992calculation}, and Townsend~\cite{townsend2023generating} for the details:
\begin{align}
    \gamma X^A_{1 1} &= \frac{3}{4\xi} - \frac{9}{40} \ln{\xi} - \frac{3}{56} \xi \ln{\xi} + A^X_{11} ; \\
    -\gamma X^A_{1 2} &= \frac{3}{4\xi} - \frac{9}{40} \ln{\xi} - \frac{3}{56} \xi \ln{\xi} - A^X_{12} ,
\end{align}
where $\xi=s-2$ is the dimension-free gap between the particles surfaces, and the constant offset are given by
\begin{align}
    A^X_{1 1} &= \frac{15}{16} + \sum^{\infty}_{m=1} \Bigg\{ 2^{-4m} f_{2m} -\frac{1}{4} -\frac{9}{40m}\nonumber \\ 
    &\phantom{ \frac{15}{16} + \sum^{\infty}_{m=1} }  + \frac{3}{56m} [(-\delta_{m1} + (m-1)(1-\delta_{m1})]^{-1}  \Bigg\}. \\
    -A^X_{1 2} &= -\frac{5}{112}+\frac{9}{20} \log_{10}{2} + \sum^{\infty}_{m=0} \Bigg\{ 2^{-4m-2} f_{2m-1} \nonumber \\ 
    &\phantom{ -\frac{5}{112}+\frac{9}{20} } \qquad -\frac{1}{4} -\frac{9}{40m} + \frac{3}{28m} (2m-3)^{-1}  \Bigg\}.
\end{align}

%%%%%%%
\section{\label{sec:appen_nn}Training a Neural Network for Many-Body Hydrodynamics}
%%%%%%%

This appendix provides details on the neural-network training procedure used to approximate many-body hydrodynamic corrections in our hybrid framework. The implementation was carried out using JAX~\cite{jax2018github} and Flax~\cite{flax2020github}, leveraging automatic differentiation and GPU acceleration for efficient optimization.

The training dataset consists of three-body mobility data computed using the original Stokesian Dynamics framework~\cite{brady1988stokesian}. For each sample, particle positions were randomly initialized within a cubic domain of size $[-2,2]$ while ensuring a minimum separation distance of $2a$ to avoid non-physical overlaps. The corresponding exact many-body mobilities were obtained from the inverse resistance matrix~\cite{durlofsky1987dynamic} and two-body reference solutions from Jeffrey and Onishi~\cite{jeffrey1984calculation}. To enhance numerical stability, the relative particle displacement
components were processed using a logarithmic transformation of the form:
\begin{align}
\tilde{r}_i = \text{sign}(r_i) \log(1 + |r_i|).
\end{align}
The dataset consists of $10^5$ samples, with an 80-20 train-test split using stratified sampling to ensure a balanced distribution of interparticle separations.

Two separate neural networks were trained to model the many-body corrections to pairwise mobilities, $\mathbf{G}_{\alpha\beta}$, and the self-mobility corrections, $\boldsymbol{\mu}_{\alpha\alpha}$. Both networks were implemented as fully connected multi-layer perceptrons (MLPs) using Flax. Each network consists of two hidden layers with 256 neurons per layer and Rectified Linear Unit (ReLU) activations. A final linear layer maps to the output dimension, ensuring the appropriate rank for the predicted mobility corrections. To enforce physically meaningful outputs, the predictions are constrained such that:
\begin{align}
\mathbf{G}_{\alpha\beta} = \mathbf{I} + \text{NN}_{G}(X), \quad \boldsymbol{\mu}_{\alpha\alpha} = \mathbf{I} + \text{NN}_{\mu}(X),
\end{align}
where $X$ represents the processed relative displacements, and $\mathbf{I}$ is the identity matrix. These constraints ensure that the learned corrections act as perturbations on the analytically known two-body mobilities.

The networks were trained using the Adam optimizer~\cite{kingma2017adammethodstochasticoptimization} with a learning rate of $10^{-3}$. The training objective was to minimize the mean squared error $\mathcal{L}_{\text{MSE}}$ between the predicted and exact many-body mobilities:
\begin{align}
\mathcal{L}_{\text{MSE}} = \frac{1}{N} \sum_{i=1}^{N} \left| \text{NN}(X_i) - Y_i \right|^2.
\end{align}
Training was performed for 500 epochs with mini-batches of 512 samples. Model checkpoints were saved using Optax~\cite{deepmind2020jax}, and final model weights were stored in a compressed format to facilitate reproducibility.

\end{document}